\documentclass[preprint,showpacs,prl,preprintnumbers,amsmath,amssymb]{revtex4}

\usepackage{graphicx}
\usepackage{dcolumn}
\usepackage{bm}

\begin{document}
\title{Magnetotransport in Single Crystal Half-Heusler Compounds}

\author{K. Ahilan$^{1)}$, M. C. Bennett$^{1)}$, M. C. Aronson$^{1)}$, N. E. Anderson$^{2}$, P. C. Canfield$^{2)}$,
E. Munoz-Sandoval$^{3),4)}$, T. Gortenmulder$^{3)}$, R.
Hendrixx$^{3)}$, and J. A. Mydosh$^{3),5)}$, }

\address{$^{1)}$ University of Michigan, Ann Arbor, MI 48109, USA\\
 $^{2)}$ Ames Laboratory, Iowa State University, Ames, IA 50011
 USA\\
 $^{3)}$ Kamerlingh Onnes Laboratory, Leiden University, Leiden, The Netherlands.\\
 $^{4)}$ Advanced Materials Department, IPICyT, Apartado Postal 3-74,78231 San Luis Potosi, S. L. P, Mexico.\\
 $^{5)}$ Max Planck Institute for Chemical Physics of Solids, Dresden, Germany.\\}
\begin{abstract}
{We present the results of electrical resistivity and Hall effect
measurements on single crystals of HfNiSn, TiPtSn, and TiNiSn.
Semiconducting behavior is observed in each case, involving the
transport of a small number of highly compensated carriers.
Magnetization measurements suggest that impurities and site
disorder create both localized magnetic moments and extended
paramagnetic states, with the susceptibility of the latter
increasing strongly with reduced temperature. The
magnetoresistance is sublinear or linear in fields ranging from
0.01 - 9 Tesla at the lowest temperatures. As the temperature
increases, the normal quadratic magnetoresistance is regained,
initially at low fields, and at the highest temperatures extending
over the complete range of fields. The origin of the vanishingly
small field scale implied by these measurements remains unknown,
presenting a challenge to existing classical and quantum
mechanical theories of magnetoresistance.}

\end{abstract}
\pacs{71.20.Lp.71.20.Nr,72.20.-i} \maketitle


One of the most exotic settings for small gap insulators is among
materials made entirely of metals, i.e. intermetallic compounds.
Nonetheless, it was demonstrated in the early 1990's that a subset
of the half-Heusler compounds, with the generic formula RNX (R,N
are transition metal elements, and X is a main group element),
display semiconducting transport while infrared absorption
experiments found semiconducting gaps with magnitudes of $\sim$100
meV or less~\cite{aliev1989,aliev1990}. In subsequent years, these
materials were found to have a substantial
thermopower~\cite{aliev1990,cook1996}, and a number of different
compositions were
explored~\cite{uher1999,hohl1998,mastronardi1999,tobola1998,
kasaya1992} in the hopes of optimizing these characteristics. All
of these experiments were carried out on polycrystalline samples,
and since annealing dramatically affected their transport
properties~\cite{aliev1990,uher1999}, it was clear that the
samples were unlikely to be either stoichiometric, or single
phase. Understanding the intrinsic behaviors of these unusual
materials has awaited the synthesis of single crystal samples,
which we report here.

Many intermetallic compounds form in the Heusler structure
RN$_{2}$X, with four interpenetrating face centered cubic
lattices, where the N elements lie at the center of the rock-salt
cube formed by the R and X elements. The half-Heuslers are a
variant on this structure, where one of the sublattices is
entirely vacant, implying that now the N atoms ideally occupy
every other rock-salt cube. This MgAgAs structure is found in
intermetallic compounds with 16-20 electrons per formula unit z,
but most often when z=18 as in the RNiSn (R=Ti, Zr, and
Hf)~\cite{dwight1974}. Of the known half-Heusler
compounds~\cite{hohl1999}, only a few are thought to be insulating
in the absence of disorder: RNiSn (R= Ti, Zr, and Hf) and RNSn
(R=Zr and Hf, N=Pd and Pt)~\cite{aliev1989}, LnPdSb (Ln=Ho,Er,
Dy)~\cite{mastronardi1999}, LnPtSb (Ln=trivalent rare
earth)~\cite{kasaya1992}, CoTiSb and CoNbSn~\cite{tobola1998}, and
NbIrSn~\cite{hohl1998}. Studies of half-Heusler systems in which
the electron count is changed systematically
~\cite{pierre1994,tobola1996, tobola1998} reveal that insulating
behavior is generally observed near z=18, while metallic and
sometimes magnetically ordered systems are found at both higher
and lower electron counts. We note that this result is based
largely on resistance measurements carried out on polycrystalline
samples, and that site disorder is potentially large in the
half-Heusler structure, even for nominally stoichiometric
compositions~\cite{aliev1989,uher1999}. For these reasons, the
exact range of intrinsic insulating behavior, is still
experimentally uncertain. Electronic structure
calculations~\cite{ogut1995} support the view that the
semiconducting behavior found in the RNiSn (R=Ti,Zr, Hf) is an
intrinsic feature, finding indirect gaps between the $\Gamma$ and
X points in each compound. The gap is determined by the strength
of the R-Sn pd hybridization, which leads to anticrossing of the
extrema of the conduction and valence bands, which have primarily
R d$_{xy}$ character. The anticrossing is further amplified
through an indirect interaction of the R states mediated by low
lying Ni states. Nonetheless, the indirect gaps in TiNiSn, ZrNiSn,
and HfNiSn are expected to be nearly identical, $\sim$ 0.5 eV.

The purposes of this paper are two fold. First, we report the
results of transport and magnetization experiments on single
crystals of TiNiSn, HfNiSn, and TiPtSn. For comparision, similar
experiments were carried out on polycrystalline samples as well.
These measurements show that the semiconducting behavior found in
polycrystalline samples is at least partly intrinsic, although
electron microscopy evidence is presented which reveals
substantial metallurgical phase separation in all polycrystalline
samples. All of the samples, both single crystal and
polycrystalline, are weakly magnetic, due to the presence of both
extended and localized moment bearing defect states. The Hall
constant is unusually small, suggesting a near balance of
electrons and holes. While our original objective was to establish
the RNiSn as intrinsic semiconductors by measuring single crystals
instead of polycrystals, we found in addition that the
magnetotransport is anomalous in this family of compounds. The
second section of this paper is devoted to a description of the
linear magnetoresistance found above a field scale which grows
linearly with temperature, and which varies only moderately among
our different samples.

Polycrystalline samples of HfNiSn were prepared by arc melting in
a high purity Ar atmosphere, followed by high temperature vacuum
annealing. An electron backscattering image of the as cast
material is shown in Fig. 1a, revealing that most of the sample is
near-stoichiometric HfNiSn, but that there are additional phases
at the boundaries of the grains. Microprobe measurements find that
these phases are roughly equal quantities of  Hf$_{5}$Sn$_{4}$ and
the Heusler compound HfNi$_{2}$Sn, together making up about 4$\%$
of the total as cast sample volume. After annealing the as cast
material in vacuum at 1000 C for 720 hours ~\cite{aliev1989}, the
grain boundary phases are much reduced in volume (Fig. 1b), and we
presume that the composition of the grains approaches the
stoichiometric level. Annealing had very similar effects on both
TiNiSn and ZrNiSn polycrystals, which were found to undergo even
more extensive phase separation. We also attempted to introduce
B,Ta,Lu,Co,Mn,U,and Ce into HfNiSn and ZrNiSn as dopants. In every
case, the dopants were segregated in the grain boundary phases,
with no measurable solubility in the HfNiSn and ZrNiSn grains. We
note that earlier efforts to dope the RNiSn also employed some of
these elements~\cite{hohl1999,uher1999,slebarski1998}, and our
results suggest that the primary effects of doping observed in
these works likely result from modifications to the conducting
properties of the grain boundary phases, and not from changes in
the RNiSn matrix itself. We also experimented with shorter anneals
at higher temperatures. Fig. 1c shows that a 72 hour anneal at
1300 C significantly decreases the volume of grain boundary
phases, which is now largely Hf$_{5}$Sn$_{4}$, although higher
levels of HfNi$_{2}$Sn and significant Sn loss from the matrix are
observed near the surface of the polycrystalline sample. Due in
part to these concerns about sample homogeneity, single crystals
of HfNiSn were prepared at the University of Amsterdam/Leiden
University ALMOS facility using the tri-arc assisted Czolchralski
method, while single crystals of TiNiSn and TiPtSn were
synthesized at Ames Laboratory from a Sn flux~\cite{canfield1992}.
The crystal structures of all HfNiSn samples, both single crystal
and polycrystal, were verified to be the MgAgAs type by powder
x-ray diffraction.

We have studied these samples using both magnetization
measurements, carried out using a Quantum Designs SQUID
magnetometer, and electrical transport measurements, which
employed both a Quantum Designs Physical Phenomena Measurement
System and homebuilt helium cryostat systems. The resistivity and
Hall effect measurements were performed at a frequency of 17 Hz,
in the conventional four-probe and five probe configurations,
respectively. Sweeps in positive and negative fields were combined
to separate the mixed Hall signal and longitudinal
magnetoresistance at each temperature. Special care was taken to
avoid heating by the measuring current, especially at the lowest
temperatures.

Sample homogeneity has a profound effect on the transport observed
in the half-Heuslers. Fig. 2 depicts the temperature dependent
resistivity for four different samples of HfNiSn: single crystal,
as cast polycrystal, and polycrystal annealed for 720 hours at
1000C (optimally annealed), and 72 hours at 1300 C. The
resistivity  increases slowly with decreasing temperature in the
as cast material and reaches a broad maximum near 120 K, before
decreasing and displaying a superconducting transition near 3.8 K.
The superconductivity is easily suppressed with modest fields, and
since it is absent in the optimally annealed and single crystal
samples we ascribe it to the grain boundary phases, or to trace
amounts of elemental Sn. The resistivity of the optimally annealed
polycrystalline sample reproduces the insulating behavior of the
single crystal, although its resistivity  is more than an order of
magnitude smaller for the entire temperature range. The 72
hour/1300C anneal has an intermediate effect, marginally
increasing the measured resistivity over that of the as-cast
sample, while rendering it insulating down to the superconducting
transition. We conclude from the data in Fig. 2 that the grain
boundary phases are highly conducting relative to the matrix, but
that once they are removed by annealing HfNiSn is intrinsically a
semiconductor, as previously claimed on the basis of experimental
evidence gathered from the polycrystalline
samples~\cite{aliev1989}.

Despite their intermetallic character, the single crystals of
HfNiSn, TiNiSn, and TiPtSn all display semiconducting behavior, as
predicted for filled d-band half-Heuslers~\cite{ogut1995}. Fig. 3a
shows the temperature dependences of the electrical resistivity of
samples of all three materials, showing the behavior typical of
n-type semiconductors. The activation plots of Fig. 3b are not
linear over the entire range of temperatures investigated. In
every case, the resistivity slowly approaches a constant value at
the lowest temperatures, but in addition there are regions of
distinctly different slopes, especially for TiNiSn. These results
indicate the presence of narrow bands of impurity/defect states in
the semiconducting gap, with excitation energies which are smaller
than the semiconducting gap itself. The highest temperature
transport gaps are very small: 26 meV for HfNiSn, 28 meV for
TiPtSn, and 79 meV for TiNiSn. The estimates for the transport
gaps in HfNiSn and TiNiSn are much smaller than the values
previously reported for polycrystals~\cite{aliev1989}, as well as
being smaller than the indirect gap of 0.5 eV predicted by
electronic structure calculations for TiNiSn and HfNiSn
~\cite{ogut1995}.

Hall effect measurements have been carried out on single crystals
of HfNiSn, TiNiSn and TiPtSn, in order to estimate the number and
type of carriers present. The data are presented in Fig. 4a. No
Hall signal was detected in single crystal or polycrystalline
HfNiSn in fields as large as 9 Tesla, and for temperatures between
1.2 K and 12 K. Our experimental accuracy consequently yields only
an lower bound for a single band carrier concentration n of
5x10$^{21}$ cm$^{-3}$, which is more than one carrier per unit
cell. The Hall constant in TiPtSn is just at the limits of our
experimental resolution, and also indicates a large carrier
concentration $\sim$ 1x10$^{21}$cm$^{3}$. In view of the
semiconducting characters of single crystal HfNiSn and TiPtSn, we
view these single band carrier concentrations as unreasonably
high. We consider it more likely that both HfNiSn and TiPtSn are
very close to perfect compensation, and perhaps may even be
semimetals. In contrast, a large Hall voltage was detected in
TiNiSn, linear in field and with a slope which indicates that the
dominant carriers are electrons. The temperature dependence of the
electron concentration n deduced from these measurements is
plotted in Fig. 4b, showing a strong increase with reduced
temperature, amounting to an increase by approximately one
electron per 365 unit cells between 40 K and the approximately
constant value of 1x10$^{19}$ cm$^{-3}$ found below 15 K. It is
significant that the low temperature electron concentration in
TiNiSn is approximately the same as the number of paramagnetic
moments inferred from the magnetization measurements, which we
discuss next.

All of the single crystal and polycrystalline samples we measured
are weakly magnetic, despite the nominally nonmagnetic character
expected for the filled-d shell half-Heusler compounds. The field
dependence of the magnetization of polycrystalline HfNiSn,
annealed for 720 hours at 1000 C, is presented in Fig. 5a, while
similar data for single crystal TiNiSn appear in Fig. 5b. In both
cases, the magnetization is strongly nonlinear, especially at low
fields and low temperatures. As the temperature is raised, the
magnetization is increasingly dominated by a diamagnetic
contribution, linear in field, which is especially evident for the
TiNiSn sample. The magnetization M divided by a constant measuring
field of 1.5 T is plotted in Fig. 6 as a function of temperature
for three different samples: polycrystalline HfNiSn, both as cast
and annealed for 720 hours at 1000 C, as well as single crystal
TiNiSn. Fig. 6 shows that the paramagnetic contribution to the
magnetization is largest at low temperatures, but is superposed on
a diamagnetic contribution. It is evident that the magnitudes of
the paramagnetic and diamagnetic components, as well as their
relative weights, vary considerably among the samples we
investigated.

Figures 5 and 6 suggest that it may be possible to separate the
two components of the magnetization by modelling the magnetization
in each sample as M(H,T) = $\chi_{o}$H+$\cal{F}$(H/T).  The first
term represents a field independent susceptibility, which we will
show combines the normal diamagnetic susceptibility of the
semiconducting host, with an additional Pauli paramagnetic
susceptibility, which varies among samples. Given the small
magnitude of the moments shown in Fig. 5, we think it unlikely
that long range magnetic order is responsible for the nonlinear
magnetization found at low temperatures. Instead, we suggest that
the second term represents the scaling behavior of isolated and
localized magnetic moments, where $\cal{F}$ is consequently
expected to be the Brillouin function.

The results of this modelling are shown in Fig. 7, which
demonstrates that the magnetization  sweeps taken in both samples
for temperatures which range from 2 K - 20 K collapse onto scaling
curves if a diamagnetic term is previously subtracted from the
data at each temperature. As indicated, the scaled and corrected
magnetizations for HfNiSN and TiNiSn are well described by J=1/2
Brillouin functions, yielding moment densities of
1.98x10$^{-3}$$\mu_{B}$ per formula unit TiNiSn and
1.55x10$^{-3}$$\mu_{B}$ and 0.144 $\mu_{B}$ per formula unit
HfNiSn for the two annealed and polycrystalline HfNiSn samples
investigated. We find that the diamagnetic susceptibilities
$\chi_{o}$ of the three systems are also very different, as Fig. 6
suggests. These susceptibilities are plotted in Fig. 8, showing
that $\mid\chi_{o}\mid$ is much larger for TiNiSn than for either
HfNiSn sample, although $\mid\chi_{o}\mid$ varies by almost an
order of magnitude between the two HfNiSn samples. While
$\chi_{o}$ approaches a constant value at low temperature in one
of the polycrystalline HfNiSn samples (filled circles), in each of
the three samples, $\mid{\chi_{o}}\mid$ increases approximately
linearly with temperature.

The pronounced temperature and sample dependences of $\chi_{o}$
which are demonstrated in Fig. 8 argue strongly for the presence
of a paramagnetic contribution to the magnetization which is
linear in field. The susceptibility of a semiconductor is
approximated by the sum of the core susceptibilities of the
constituent atoms: -4.8x10$^{-5}$ emu/g for HfNiSn and
-3.7x10$^{-5}$ for TiNiSn\cite{selwood1956}, and is consequently
independent of temperature and invariant among samples of the same
nominal composition. The values we find for $\chi_{o}$ in HfNiSn
and TiNiSn do not approach these limits on the temperature range
of our experiment, suggesting the presence of an additional
positive magnetization, linear in field and decreasing in
magnitude with increasing temperature. The variation of $\chi_{o}$
between the two HfNiSn samples suggests that this inferred
positive susceptibility has an extrinsic origin, consistent with
the small magnitude of the local moments found in each sample, and
with the variations in moment concentrations found among different
samples.

We conclude that there are two magnetic entities present with
different relative weights in each of our half-Heusler samples,
superposed on the diamagnetic response of the semiconducting host.
The first entities are localized magnetic moments, perhaps
generated by the inevitable site disorder characteristic of the
half-Heusler structure. Still, the level of this putative disorder
can be quite low, corresponding to only one defect per 7000 unit
cells in our least magnetic HfNiSn sample. Secondly, we propose
that there are also extended and metallic states present which are
responsible for the inferred paramagnetic susceptibility. In
agreement with Hall effect measurements on TiNiSn, the number of
these metallic states grows with decreasing temperature. It is
interesting that the electron concentration in $\mbox{TiNiSn}$ is
similar to the concentration of localized magnetic moments,
suggestive that both have the same origin, presumably defects or
impurities. Further, the data suggest that the electron
contributed by each impurity or defect in TiNiSn ultimately
becomes delocalized as T$\rightarrow$0, which is, intriguingly,
just what is expected if the moment bearing impurities are
compensated by the Kondo effect.

Perhaps the most striking property of the nonmetallic half-Heusler
compounds is their magnetoresistance, which is linear or even
sublinear in field at the lowest temperatures. The anomalous
magnetoresistance has been observed in every half-Heusler compound
we have measured, both single crystals and polycrystals, both as
cast and annealed. This is demonstrated in Fig. 9 which presents
the 2 K magnetoresistances of HfNiSn, TiPtSn, and TiNiSn single
crystals, and HfNiSn polycrystal, both as cast and annealed for
720 hours at 1000C. The magnetoresistances of the single crystal
samples are markedly sublinear, while those of the polycrystalline
samples are more highly linear. In neither case is there any
suggestion of a quadratic magnetoresistance, down to fields as
small as 1000 Oe. We note that trace superconductivity in the
polycrystalline samples prohibits measurements to fields less than
$~\sim$ 500 Oe. The magnitude of the magnetoresistance is rather
modest, amounting to a few percent in each of the crystalline
samples. It approaches 10$\%$ at 9 T in the as-cast
polycrystalline sample, although annealing reduces the
magnetoresistance to the level observed in the single crystals.
This may result either from a compactification of the internal
structure of the polycrystals, since annealing removes conducting
grain boundaries, or from rendering the HfNiSn grains more nearly
stoichiometric, as in the single crystal.

The low field magnetoresistance of all samples ultimately becomes
quadratic in field if the temperature is increased sufficiently.
The longitudinal magnetoresistance of single crystal HfNiSn is
presented at several temperatures in Fig. 10, with an expanded low
field region in the inset. At 2 K, the magnetoresistance is
sublinear over our entire field range, and is never quadratic,
even at the lowest fields. At 4 K, the magnetoresistance is linear
over virtually the entire range of fields studied, with little
trace of the upward curvature which emerges at the lowest fields
at 6 K. Raising the temperature still furthur yields a range of
fields for which quadratic magnetoresistance is found. This is
demonstrated in Fig. 11, where the magnetoresistance of single
crystal HfNiSn is plotted as a function of H$^{2}$. At 30 K and
above, the magnetoresistance is quadratic in field for the entire
range of fields studied, 0.01 -9 T. As the temperature is lowered,
clear departures from $\Delta\rho$/$\rho$ $\propto$ H$^{2}$ are
observed, below fields H$^{*}$ which become progressively smaller
as the temperature is reduced. This analysis has been repeated on
all of our samples, and the results are summarized in Fig. 12. It
is evident that the range of fields H$\leq$H$^{*}$ and
temperatures for which $\Delta\rho$/$\rho$ $\propto$ H$^{2}$
varies considerably among the different samples. For instance, the
magnetoresistance of the annealed sample of HfNiSn(open circles,
Fig. 12) is quadratic in fields up to 9 T for all temperatures
above 15 K, while clear departures are still evident at
temperatures as large as 100 K for an as cast polycrystalline
sample of HfNiSn(open squares, Fig. 12).

Fig. 12 may be viewed as an organizational scheme for our
magnetoresistance observations, analagous to a phase diagram,
where H$^\ast$(T) represents a crossover from a regime at low
fields where the magnetoresistance is quadratic in field, to a
regime at high fields where the magnetoresistance is linear or
even sublinear in field. In agreement with the discussion of Fig.
9, the linear magnetoresistance is seen for the widest range of
temperatures in the as cast polycrystalline samples, while
annealing confines the linear magnetoresistance to the lowest
temperatures and highest fields. H$^{*}$(T) for the single
crystals is similar to that of the most homogeneous
polycrystalline samples, which have been annealed for the longest
times.

The observation of a linear or sublinear magnetoresistance is
anomalous, as the magnetoresistance typically depends on the field
magnitude, yielding a magnetoresistance which is even and usually
quadratic in the applied field. Nonetheless, there are a few
circumstances in which a linear magnetoresistance is expected. A
linear magnetoresistance can occur in inhomogeneous materials,
resulting from the accidental admixture of the Hall
signal\cite{herring1960}. We rule out this possibility for the
half-Heuslers, as we observe linear magnetoresistance in single
crystals, and further since the Hall signal is immeasurably small,
especially in HfNiSn and TiPtSn. Similarly, a variety of different
mechanisms can lead to linear magnetoresistance in systems with
large carrier concentrations\cite{budko1998,young2003}, but in
most cases their application to the minimally magnetic and
essentially semiconducting half-Heuslers seems tenuous. Finally,
we note that there is no evidence for either magnetic or
structural transitions, which might provide an internal symmetry
breaking field, superseding the applied field.

Linear magnetoresistance is expected in metals with closed Fermi
surface orbits at intermediate fields, in a crossover regime
between the quadratic magnetoresistance expected at low fields,
and saturation at high fields. The range of fields over which
linear magnetoresistance can be observed is potentially quite
extensive for materials with a high degree of
compensation \cite{pippard 1989}. However, this linear crossover
regime emerges when the product of the cyclotron frequency
$\omega_{c}$ and the scattering time $\tau$ exceeds one,
$\omega_{c}\tau\geq$1. For TiNiSn at 2 K, where n=1.3x10$^{19}$
and $\rho$=18$\Omega$-cm, $\omega_{c}\tau$=1 for an applied field
of 3.7x10$^{5}$ Tesla. Indeed, this crossover field could be
considerably larger if we consider that the Hall measurements
provide only a lower bound on the total number of carriers, which
may be electrons as well as holes. Conversely, if we assume that
this crossover field is no more than 100 Oe, consistent with our
lowest temperature magnetoresistance results on single crystal
HfNiSn and TiPtSn, we would require a total carrier concentration
of $\sim$3x10$^{13}$cm$^{-3}$. It is difficult to believe that the
numbers of electrons and holes present in both our single crystal
and polycrystalline samples are balanced at the level of one
carrier per 10${^8}$ unit cells, as the absence of a Hall signal
suggests. We are forced to conclude that in each of our samples,
linear or sublinear magnetoresistance is observed for a wide range
of fields which are comfortably in the low field limit
$\omega_{c}\tau\ll$1.

Linear magnetoresistance is also predicted theoretically for a low
density electron gas, in which only the lowest Landau level is
occupied~\cite{abrikosov1969,abrikosov1998,abrikosov2000}.  The
two requirements for realizing this quantum magnetoresistance are
that H$\geq$($\hbar$cn)$^{2/3}$, and that
T$\ll$eH$\hbar$/m$^{\ast}$c. Consequently, the linear
magnetoresistance should only be observed in TiNiSn at fields
greater than 36 Tesla, and for temperatures T$\ll$1.35 (K/T)H. It
is possible that the latter condition is satisfied in the
half-Heusler compounds, since as Fig. 11 demonstrates, the linear
magnetoresistance is only observed above a field H$^{\ast}$ which
increases linearly with temperature. In this view, the different
slopes found in the plot of H$^{\ast}$(T) for the different
systems (Fig. 11) could plausibly be the result of variations in
magnitude of the effective mass between the least massive
(m$^{\ast}$/m=0.09 in as cast HfNiSn 99106) and the heaviest
(m$^{\ast}$/m=0.58 in single crystal TiPtSn). Nonetheless, our
observation of linear magnetoresistance in TiNiSn in fields as
small as 0.01 T, where the criterion H$\geq$($\hbar$cn)$^{2/3}$ is
dramatically violated is problematic for the quantum
magnetoresistance theory as well. One possibility is that only a
tiny fraction of the carriers detected by the Hall effect in
TiNiSn, one carrier in $\sim$ 2x10$^{5}$, actually participates in
the quantum magnetoresistance. We note that a similar explanation
has been proposed \cite{abrikosov1998} as an explanation for the
linear magnetoresistance observed in doped silver chalcogenide
compounds~\cite{husmann2002,lee2002,xu1997,schnyders2000}. Here,
spatial separation of conducting and insulating regions was
invoked to achieve the needed reduction in effective carrier
concentration, a scenario which is inapplicable to the single
crystal half-Heuslers. The central dilemma lies with explaining
the unexpected persistance of the linear and sublinear
magnetoresistance to extremely low fields, particularly at low
temperatures. If the linear magnetoresistance observed in the
half-Heuslers derives from the scattering of fully quantum
mechanical states, then our measurements suggest a less
restrictive set of necessary conditions than has been
theoretically proposed.

It is unclear whether the linear magnetoresistances found in the
half-Heuslers and the silver chalcogenides arise from the same
fundamental mechanism. Certainly there are general resemblences
between the doped silver chalcogenides and the semiconducting
half-Heuslers, as both systems are nonmagnetic semiconductors,
with small or vanishing values of the Hall constant. We note that
the single crystal half-Heuslers generally have larger
resistivities, and as far as is known, electron concentrations
which are similar to those in optimally doped
Ag$_{2-\delta}$Te\cite{lee2002}. The progression of the
magnetoresistance from sublinear, to linear, and ultimately to
quadratic field dependencies is achieved by increasing temperature
and consequently decreasing electron concentration in the
half-Heuslers. In contrast, temperature has a very modest effect
on the linear magnetoresistance found in the doped silver
chalcogenides\cite{xu1997}, although the same progression of
field dependencies is achieved in this system by using high
pressures to drive the Hall constant through zero\cite{lee2002}.
A similar sensitivity to carrier density is apparently absent in
the half-Heuslers, since the magnetoresistances of single crystal
HfNiSn, TiPtSn, and TiNiSn are very similar in magnitude, despite
their very different electron concentrations.  Finally, we note
that the magnetoresistances of the single crystal half-Heuslers
are also much smaller than those of the silver chalcogenides, and
the uniformly larger magnetoresistance found in as-cast
polycrystalline half-Heusler samples may support the suggestion
that compositional inhomogeneities, present only in the
polycrystalline half-Heusler samples, are required to obtain
large, linear magnetoresistances\cite{abrikosov1998,lee2002}.

We have established here that single crystals of several
half-Heusler compounds, selected to have a total of 18 valence
electrons, are small gap insulators, in  agreement with both
electronic structure calculations and previous measurements on
polycrystalline samples. Resistivity measurements show that
HfNiSn, TiPtSn, and TiNiSn single crystals are all semiconducting,
while magnetization measurements argue that both localized
magnetic and extended states are introduced with impurities or
site interchange disorder. Hall effect measurements indicate that
the number of electrons and holes is closely balanced in HfNiSn
and TiPtSn, although there is a pronounced excess of electrons in
TiNiSn. Both the corresponding electron concentration and the
inferred Pauli susceptibility in TiNiSn increase with decreasing
temperature, which may signal the return via the Kondo effect of
electrons to the Fermi surface which were formerly localized at
high temperatures in magnetic impurity states.

Our results deepen the mystery surrounding the origin of linear
magnetoresistance in materials with small carrier densities.
Neither the classical nor quantum mechanical theories of the
magnetoresistance can countenance the persistance of the linear
magnetoresistance to fields as low as 0.01 T, given the mobility
and carrier densities characteristic of both half-Heuslers and the
doped silver chalcogenides. The observation of the linear
magnetoresistance in single crystals of the half-Heuslers, which
are presumed to be compositionally homogeneous on the shortest
length scales, removes a degree of freedom present in the doped
silver chalcogenides. Barring another mechanism which dramatically
limits the number of electrons responsible for the linear
magnetoresistance in the half-Heuslers, a new explanation for the
origin of the miniscule field scale in these unusual materials
must be sought.

We acknowledge stimulating discussions with C. M. Varma, as well
as the assistance of S. Ramakrishnan in early stages of the sample
synthesis. MCA is grateful to the Dutch NWO and FOM for partial
support during a sabbatical visit to Leiden University, where this
project was initiated. Work at the University of Michigan was
performed under the auspices of the U. S. Department of Energy
under grant DE-FG02-94ER45526. MCB acknowledges partial financial
support from NHMFL-Los Alamos. EMS acknowledges partial financial
support from CONACYT(Mexico) through Grant No. 39643-F. This
manuscript has been authored by Iowa State University of Science
and Technology under Contract No. W-7405-ENG-82 with the U. S.
Department of Energy. The United States Government retains and the
publisher, by accepting the article for publication, acknowledges
that the United States Government retains a nonexclusive, paid-up,
irrevocable, world-wide license to publish or reproduce the
published form of this manuscript, or allow others to do so, for
United States Government purposes.

\begin{figure*}[b]
\caption[]{Electron backscattering images of as cast
polycrystalline HfNiSn(a), and after annealing for 720 hours at
1000C (b) and for 72 hours at 1300 C(c). The white lines in each
panel indicate a length of 100 microns.} \label{backscattering}
\end{figure*}
\begin{figure*}[b]
\caption[]{The temperature dependent resistivity $\rho$ for as
cast polycrystalline HfNiSn (filled circles), annealed
polycrystalline HfNiSn (open circles: 720 hours at 1000 C; open
squares: 72 hours at 1300 C), and for a single crystal of HfNiSn
(filled squares). Note that the resistivity of the HfNiSn single
crystal has been divided by 25 for comparison to the
polycrystalline samples.} \label{hfnisn_rho_t_samples}
\end{figure*}
\begin{figure*}[b]
\caption[]{(a): Temperature dependent resistivities of single
crystal TiPtSn (filled circles), TiNiSn (stars), and HfNiSn
(filled diamonds). (b) Activation plots for the same data. Dashed
lines are activation fits over the indicated temperature ranges.}
\label{rho_t_single_crystals}
\end{figure*}
\begin{figure*}[b]
\caption[]{(a): Hall constants R$_{H}$ and (b) the corresponding
single band electron concentrations n for TiNiSn and TiPtSn single
crystals. Solid lines indicate upper limit for R$_{H}$ and lower
limit for n for single crystal HfNiSn.} \label{hall_effect}
\end{figure*}
\begin{figure*}[b] \caption[]{(a)Magnetization M(H) of a HfNiSn polycrystal annealed at 1000C for 720 hours,
and a TiNiSn single crystal(b). Note the large diamagnetic
magnetization in (b), and the relatively much smaller diamagnetic
contribution in (a), evident only in the 2 K M(H) curve above 3
T.} \label{hfnisn_tinisn_m_h}
\end{figure*}
\begin{figure*}[b] \caption[]{Temperature dependence of the magnetization M divided by a 1.5 Tesla measuring field H for polycrystalline HfNiSn,
both as cast and annealed for 720 hours at 1000 C, and for a
single crystal of TiNiSn.} \label{chi_1p5t_allsamples_vs_t}
\end{figure*}
\begin{figure*}[b] \caption[]{Magnetization sweeps, corrected for a linear diamagnetic susceptibility,
collapse onto scaling curves as a function of g$\mu_{B}$H/k$_{B}$T, for both polycrystalline
HfNiSn (annealed at 1000 C for 720 hours) and single crystal TiNiSn.  The solid lines are
J=1/2 Brillouin functions.} \label{brillo_fits}
\end{figure*}
\begin{figure*}[b]
\caption[]{(a): The temperature dependence of the diamagnetic
susceptibility $\chi_{o}$ for annealed samples of polycrystalline
HfNiSn (filled circles and open squares) and single crystal TiNiSn
(open circles) Solid lines are power law fits, demonstrated in a
double logarithmic plot in (b). The slopes of the power law fits
to $\mid\chi_{o}\mid$ are 1.0 for one HfNiSn (filled circles) and
TiNiSn, and 1.8 for the other HfNiSn polycrystalline sample(open
squares).} \label{chi_dia}
\end{figure*}
\begin{figure*}[b]
\caption[]{The magnetoresistances of single crystals of
HfNiSn(HfNiSn(3)), TiPtSn, and TiNiSn at 2 K. For comparison, data
are shown for polycrystalline samples of HfNiSn, both as-cast
(HfNiSn(1)) and annealed for 720 hours at 1000 C (HfNiSn(2)). The
TiNiSn ,TiPtSn, and polycrystalline HfNiSn data have all been
offset for comparison.} \label{single_crystal_mr}
\end{figure*}
\begin{figure*}[b] \caption[]{The magnetoresistance $\Delta\rho$/$\rho$ for single crystal HfNiSn at
different fixed temperatures. Inset shows the same data for an
expanded range of low fields. The 2 K and 4 K data in the inset
have been offset vertically for comparison.}
\label{99107_drho_h_var_t}
\end{figure*}
\begin{figure*}[b] \caption[]{The magnetoresistance of a single crystal of HfNiSn plotted as a function of
the field squared at different temperatures. Vertical arrows
indicate the highest field H$^{\ast}$ at which quadratic field
dependences are observed, and the dashed lines are linear fits to
the data having H$\leq$ H$^{\ast}$.} \label{drho_hsquared}
\end{figure*}
\begin{figure*}[b] \caption[]{H$^\ast$(T) defines a crossover from quadratic magnetoresistances
at low fields to linear or sublinear magnetoresistance at high
fields. H$^\ast$ varies substantially among the different samples:
Single crystal HfNiSn (filled circles), two different annealed
polycrystalline HfNiSn samples (open circles, filled squares), and
the corresponding two different as cast polycrystalline HfNiSn (
filled diamonds:,open squares), and single crystals of TiNiSn
(filled triangles) and TiPtSn (open triangles).}
\end{figure*}
\label{phase_diagram}
\end{document}